# Strong Damping-Like Torques in Wafer-Scale MoTe$_2$ Grown by MOCVD

Stasiu T. Chyczewski[1#], Hanwool Lee[1#], Shuchen Li[2#] Marwan Eladl[1], Jun-Fei Zheng[3], Axel Hoffmann[2] and Wenjuan Zhu[1*]

[1]Department of Electrical and Computer Engineering, University of Illinois at Urbana-Champaign, Urbana, Illinois 61801, USA

[2]Department of Materials Science and Engineering, University of Illinois at Urbana-Champaign, Urbana, Illinois 61801, USA

[3]Entegris Inc. Danbury, CT 06810

## Abstract

The scalable synthesis of strong spin orbit coupling (SOC) materials such as 1T' phase MoTe$_2$ is crucial for spintronics development. Here, we demonstrate wafer-scale growth of 1T' MoTe$_2$ using metal-organic chemical vapor deposition (MOCVD) with sputtered Mo and $(C_4H_9)_2$Te. The synthesized films show uniform coverage across the entire sample surface. By adjusting the growth parameters, a synthesis process capable of producing 1T'/2H MoTe$_2$ mixed phase films was achieved. Notably, the developed process is compatible with back-end-of-line (BEOL) applications. The strong spin-orbit coupling of the grown 1T' MoTe$_2$ films was demonstrated through spin torque ferromagnetic resonance (ST-FMR) measurements conducted on a 1T' MoTe$_2$/permalloy bilayer RF waveguide. These measurements revealed a significant damping-like torque in the wafer-scale 1T' MoTe$_2$ film and indicated high spin-charge conversion efficiency. The BEOL compatible process and potent spin orbit torque demonstrate promise in advanced device applications.

## Introduction & Background

Transition metal dichalcogenides (TMDs) have become one of the most exciting candidates for future electronic materials owing to desirable attributes such as high carrier mobilities, tunable electronic properties, and large spin-orbit interactions [1, 2]. More recently, $MoTe_2$ has garnered interest for both its electronic and spintronic properties [3-5]. It has several stable phases, most notably the semi-conducting 2H phase and semi-metallic 1T' phase [2, 6-9]. Like many other semiconducting TMDs, the 2H phase has been studied for electronic and optical devices exploiting novelties such as a layer-dependent bandgap, broadband photosensitivity, and third harmonic optical response [10, 11]. The 1T' phase on the other hand has attracted attention for its spintronic properties [4, 5, 12]. Owing to the strong spin orbit coupling in the compound and low symmetry, 1T' $MoTe_2$ has great potential in efficient spin orbit torque (SOT) device design.

Despite $MoTe_2$'s promise in advanced devices, there remain significant challenges to its synthesis. The growth of single crystal $MoTe_2$ films on $SiO_2$ substrates is difficult to accomplish, with most chemical vapor depositions (CVD) and metal organic chemical vapor deposition (MOCVD) growths producing polycrystalline films [2]. Many methods have been explored in existing literature to produce quality films (for example, tellurizing $MoO_2$ crystals or molybdenum films), though the consistent production of quality wafer scale growths remains difficult [9, 13]. Additionally, while the small energy difference between the different $MoTe_2$ phases makes it interesting for phase changes devices, it makes it challenging to consistently control whether pure 1T' or 2H phases are grown [9, 14]. The high reaction temperature required to synthesize films with traditional precursors (~600-700 degrees Celsius) also makes most growth recipes incompatible with conventional silicon back-end-of-line (BEOL) processes. Ideally, a low temperature process capable of producing high quality films can be realized.

In this work, we demonstrate that a novel tellurium precursor combined with a sputtered molybdenum film can be used to obtain wafer scale $MoTe_2$. We found that through a temperature range of 400 to 600 degrees Celsius we could reliably grow wafer scale films of $MoTe_2$ of uniform thickness. By changing the temperature, the phase could be varied from pure 1T' to mixed 1T'-2H. We also use spin torque ferromagnetic resonance (ST-FMR) measurements to probe the potential of the grown 1T' films in SOT devices and find strong conventional damping like torques in a wafer scale MoTe$_2$ film grown via chemical deposition. These results indicate the potential of MOCVD to produce BEOL compatible MoTe$_2$ films suitable for spintronic device design.

## Results & Discussion

1T' MoTe$_2$ was synthesized from a sputtered Mo film and di-tert-butyl-telluride [(C$_4$H$_9$)$_2$Te; henceforth abbreviated as DTBT] on a SiO$_2$/Si wafer. A 285 nm SiO$_2$/Si wafer piece was cleaned using acetone, isopropyl alcohol, and deionized water. A 3 nm Mo film was then deposited on the substrate by sputtering. Growth was carried out using an MOCVD tube furnace. As the molybdenum film wasn't exposed to high temperatures between deposition and loading into the furnace, significant oxidation is not expected. Once the sample was loaded, the furnace tube was pumped down until it reached approximately 15 mTorr. After reaching the base pressure, argon was injected at 100 sccm. The bubbler containing DTBT was cooled down to 0 °C using a homemade cooling apparatus, and a carrier Ar gas flow of 3 sccm was used for delivering the DTBT through the bubbler. Reaction temperatures between 400 °C and 600 °C were used for the synthesis, and the ramp-up time was 12 minutes for all growths. After the high-temperature reaction step, the chamber was allowed to naturally cool to room temperature without the DTBT precursor flow.

The growth process produced a large-scale, uniform 1T' MoTe$_2$ film that covered the entire surface of the sample. The film showed a uniform blue color as depicted in the optical image in Fig. 1a. Raman spectra were acquired using 532 nm confocal setup as shown in Fig. 1b. The spectra displayed distinct peaks corresponding to the A$_u$ mode at approximately 107 cm$^{-1}$, the A$_g$ mode at approximately 127 cm$^{-1}$, a prominent peak for the B$_g$ mode at around 163 cm$^{-1}$, and an additional A$_g$ mode peak at approximately 257 cm$^{-1}$. These peak positions align well with previously reported values for 1T' MoTe$_2$ [11, 15, 16]. The uniformity of the 1T' MoTe$_2$ film is also shown through spatial Raman mapping (Fig. 1c-1f). The mapping results for the aforementioned peaks confirm that the surface of the sample was fully covered by the synthesized 1T' MoTe$_2$, with consistent Raman peak intensities.

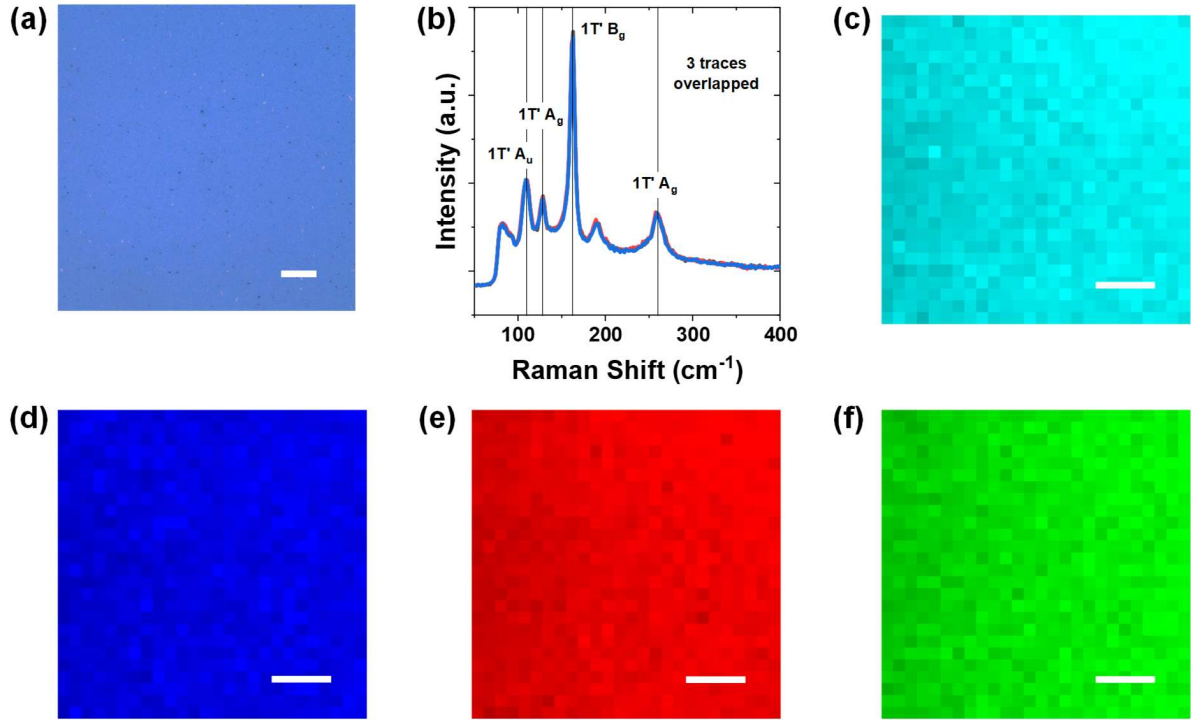

Figure 1. (a) Optical image 1T' MoTe$_2$ grown from 3 nm Mo at 500 °C with 10 μm scale bar. (b) Raman spectra from the sample. Raman mapping for 1T' MoTe$_2$ (c) 108 cm$^{-1}$ A$_u$ peak, (d) 127 cm$^{-1}$ A$_g$ peak, (e) 163 cm$^{-1}$ B$_g$ peak, and (f) 257 cm$^{-1}$ A$_g$ peak with 10 μm scale bar.

1T' MoTe$_2$ films were grown using 1 nm and 3 nm pre-sputtered Mo films. The surface characteristics and thickness of the grown films were measured using atomic force microscopy (AFM). Figure 2a and 2b show the AFM scans of 1T' phase MoTe$_2$ grown from 3 nm and 1 nm Mo films, respectively. The MoTe$_2$ film grown from the 3 nm sputtered Mo sample showed a surface roughness of 1.027 nm, while the 1 nm Mo sputtered sample showed a surface roughness of 1.008 nm. We attribute the large roughness to inhomogeneity of the precursor Mo films combined with expansion of the films during tellurization. To measure the thickness, the Mo films were patterned using photolithography prior to the growth process, allowing the synthesis of rectangular 1T' MoTe$_2$ micro strips. After the growth process, the thickness of the 1T' MoTe$_2$ flake synthesized from the 3 nm Mo layer was approximately 10 nm, while the 1 nm Mo layer sample had a thickness of around 4 nm, as shown in Fig. 2c and 2d, respectively. The film thickness increased by approximately three times after the growth, which is consistent with the structure of MoTe$_2$. The observed thickness indicates that the Mo layer was completely tellurized during the growth process, with one atomic layer of Mo sandwiched between two atomic layers of Te.

Additionally, the large-scale synthesis of 1T' MoTe$_2$ was confirmed through energy dispersive spectroscopy (EDS). EDS mapping was conducted over a large area of several square millimeters, showing the presence of both Mo and Te at every position (Fig. 2e-2f). The measured spectrum over the entire area yielded an atomic ratio of Mo 33.4 % and Te 66.6 % (~1 % error), indicating the high-quality of the synthesized 1T' MoTe$_2$ film.

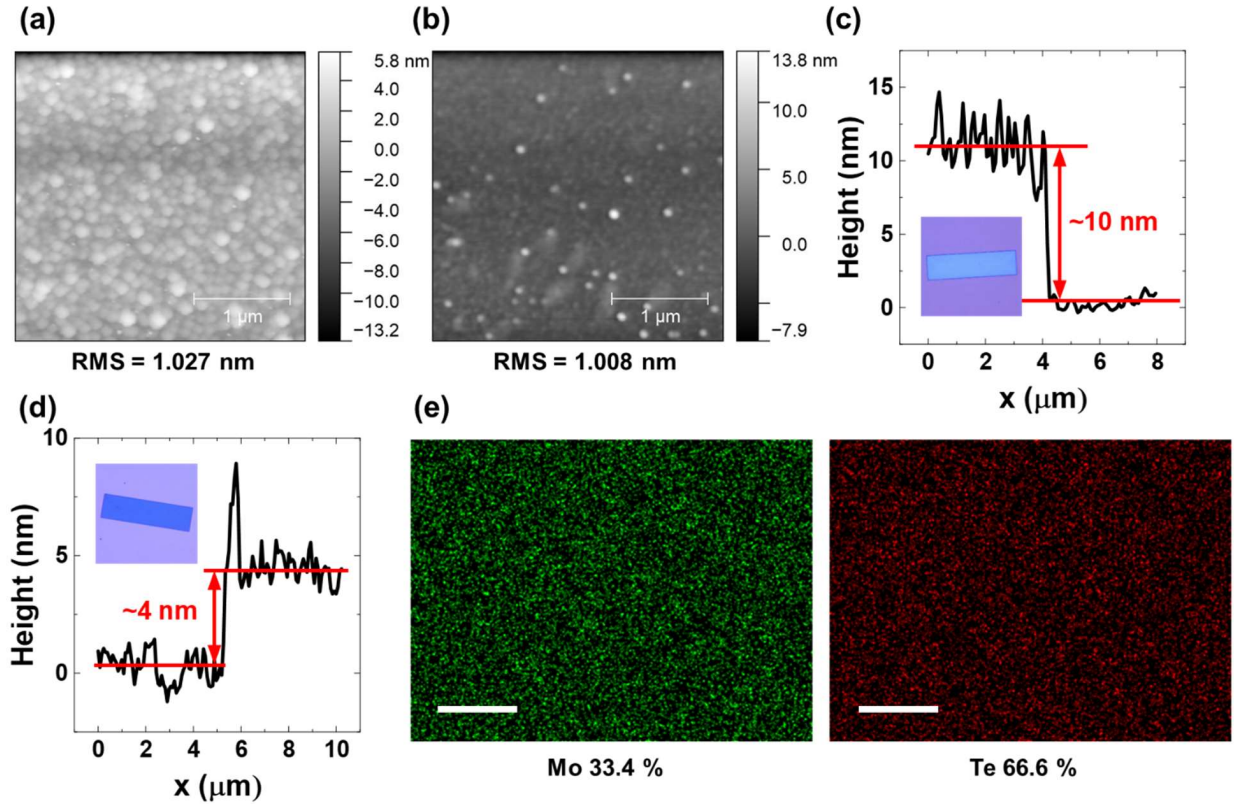

Figure 2. AFM image of 1T' MoTe$_2$ grown from (a) 3nm Mo and (b) 1nm Mo. Thickness of 1T' MoTe$_2$ grown from (c) 3nm Mo and (d) 1nm Mo. Inset shows the optical image of the film used for AFM measurement. (e) EDS mapping of 1T' MoTe$_2$ grown from 3nm Mo with 100 μm scale bar, showing that Mo to Te ratio is approximately 1:2. Reported one sigma error is ~1 %.

The growth process of 1T' MoTe$_2$ was tested at various temperatures. The Raman spectra of the 1T' MoTe$_2$ synthesized at 400 °C and 600 °C are shown in Fig. 3a. The peaks corresponding to the $A_g$, $A_u$, and $B_g$ modes in 1T' MoTe$_2$ can be clearly observed at identical positions in both cases. This indicates successful synthesis of 1T' MoTe$_2$ at both temperatures. These results suggest that our growth setup, using a sputtered Mo film and DTBT as precursors, offers a wide temperature window for successful 1T' MoTe$_2$ growth conditions. Notably, the successful growth at 400 °C shows that this process is compatible with back-end-of-line processes. We also discovered that the DTBT cannot be used with hydrogen gas. When hydrogen was introduced along with the sample and DTBT, a large-scale Te film was synthesized after growth. This was confirmed by the Raman spectra, which identified the $A_1$ mode peak of Te at around 122 cm$^{-1}$ and the $E_1$ mode peak of Te at around 142 cm$^{-1}$ [17]. Additionally, when the sample was placed upstream of the furnace, a large-scale 1T'/2H mixed-phase MoTe$_2$ film was obtained, as shown in Fig. 3b. The Raman spectra exhibited all the peaks of 1T' MoTe$_2$, along with a distinct $E_{2g}^1$ mode peak at around 234 cm$^{-1}$, confirming the presence of a mixed 1T' and 2H phase MoTe$_2$ in the obtained film. Furthermore, we discovered that annealing the film at 400 °C after growth can improve its quality. Since DTBT was used to tellurize the Mo film, Te particles were frequently observed on the sample surface after the growth process. The Raman spectra measured from a typical film synthesized without an additional annealing process showed typical 1T' MoTe$_2$ peaks along with Te peaks, and optical images reveal micrometer-scale white Te particles. To address this issue, the sample was placed in the tube furnace and annealed at 400 °C for 1 hour with 100 sccm Ar flow. After annealing, the Te peaks in the Raman spectra disappeared, and the white Te particles on the surface were no longer present. We believe that 400 °C is approximately the sublimation temperature of Te under low chamber pressure, enabling complete removal of Te without significantly impacting the quality of the MoTe$_2$ film. Note that this annealing temperature is also compatible with back-end-of-line processes.

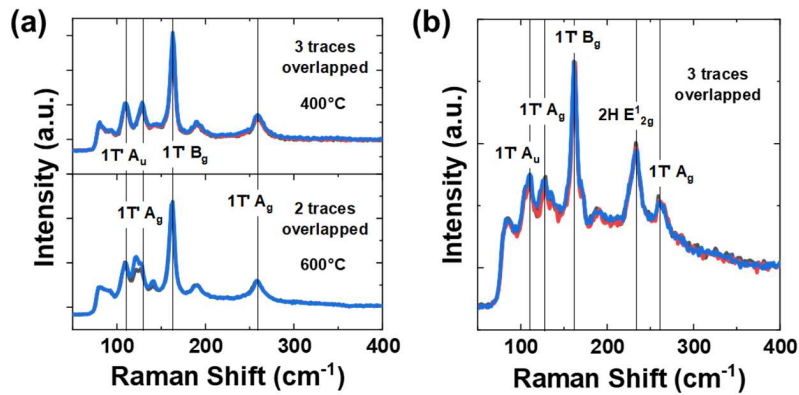

Figure 3. (a) Raman spectra of the 1T' MoTe$_2$ grown at 400 °C and 600 °C, showing wide temperature window of the successful growth condition, and back-end-of-line compatibility. (b) Raman spectra of the 1T' and 2H mixed phase MoTe$_2$ grown from the upstream of the furnace.

Magneto-transport of the MoTe$_2$ grown by MOCVD was systematically studied at various temperature and magnetic field conditions. Magneto-transport measurements offer a convenient means of probing basic properties such as carrier concentration and any significant magnetic impurities. They can also reveal exotic properties in the films (such as a very large magnetoresistance) linked to its semi-metallic nature [18]. Both longitudinal and Hall resistances were measured using the van der Pauw (vdP) technique, exploiting the wafer scale nature of the film. Temperature dependent resistivity measurements showed increasing sample resistivity with decreasing temperature (Fig. 4a). While such behavior is usually indicative of a semiconducting material, past experiments on few-layer MoTe$_2$ crystals have demonstrated similar trends [19]. The carrier concentration as extracted from these measurements was found to decrease with lower temperature, as shown in Fig. 4b, which is one of the key factors for the increasing longitudinal resistance at low temperatures. In all our samples, the Hall response was found to be linear as is expected for a non-magnetic conductor (Fig. 4c). Longitudinal magnetoresistance measurements showed a relatively small MR of ~ 0.1 %/Tesla even at very low temperatures, as shown in Fig. 4d. This is far lower than the enormous MR shown in single crystal MoTe$_2$ flakes and can possibly be explained by the polycrystalline nature of synthesized MoTe$_2$ film or oxidation of the film.

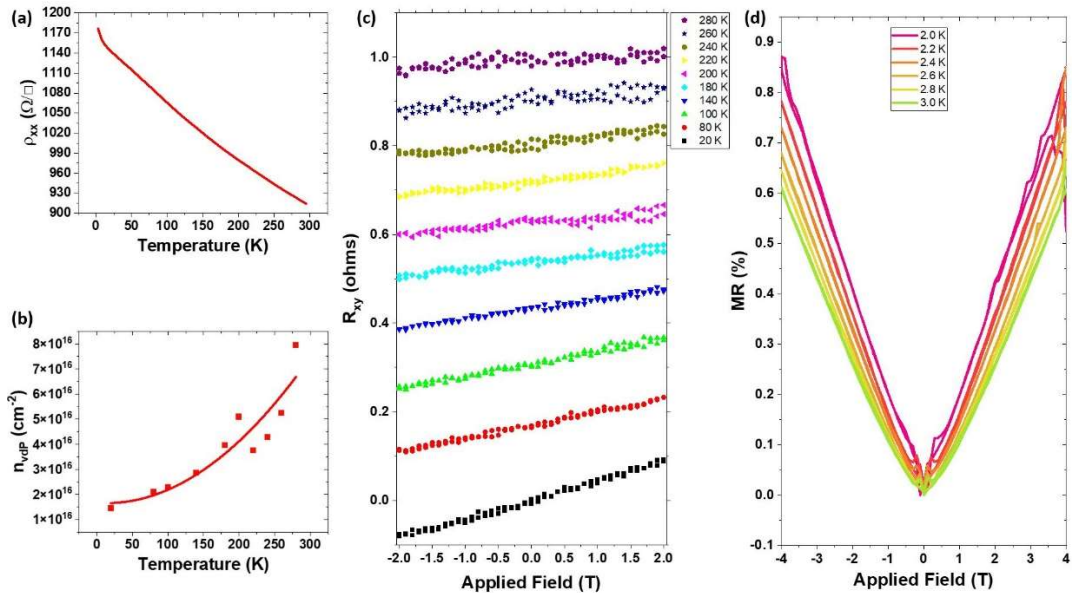

Figure 4: Magneto-transport of MOCVD MoTe$_2$. (a) Longitudinal resistance as a function of temperature. (b) Carrier density extracted from Hall resistance as a function of temperature. (c) Longitudinal magnetoresistance as a function of applied magnetic field measured at various temperatures. (d) Hall resistance as a function of applied magnetic field measured at various temperatures.

Our films are expected to exhibit strong spin-charge conversion owing to the large spin orbit coupling (SOC) that exists in MoTe₂. Devices utilizing spin orbit torques to switch magnetic layers can benefit from high speeds and magnets' intrinsic non-volatility [20]. Torques are generated via spin-charge conversion that results in spin accumulation at an interface. This is most often accomplished with the spin Hall effect (SHE). Heavy metals (e.g., platinum and tantalum) and TMDs with strong SOC in the bulk crystal (e.g., WTe₂) exhibit a potent spin Hall effect (SHE) which can be exploited in the electrical control of magnetism [21-23]. In the mono/few layer limit where the bulk SHE is not possible, spin accumulation can also result from the Rashba-Edelstein effect [24, 25]. Past work with exfoliated 1T' MoTe₂ flakes has shown that it can be used to flip adjacent magnetic layers via the SOT and displays strong non-conventional torques (i.e., accumulated spins can point in multiple directions relative to charge current direction) [5, 12].

To probe the potential of our synthesized film in spin-orbit torque applications, a coplanar RF waveguide consisting of an MoTe₂(10 nm)/permalloy(Ni₈₀Fe₂₀; 10 nm) bilayer was fabricated. Permalloy is first deposited on the entire film via sputtering with the bilayer channel subsequently defined using ion milling. The gold (100 nm) electrodes are then defined using photolithography with excess metal subsequently lifted off. A 3 nm thick layer of MgO is deposited to protect the sample. The completed structure is illustrated in Fig. 5a. Using spin torque ferromagnetic resonance (ST-FMR) measurements with an applied in-plane field to define the magnetization direction of the permalloy layer, we could assess the strength and nature of the torques present in the structure by probing the torque-modulated permalloy magnetization dynamics from the homodyne voltage signals generated by the homodyne mixing of the applied 9 GHz RF current and permalloy anisotropic magnetoresistance [23, 26, 27]. Figure 5b shows the measured ST-FMR voltage signal, with symmetric (S) and anti-symmetric (A) components extracted using the relation $V_{mix} = S \frac{\Delta H^2}{(H_{ext} - H_0)^2 + \Delta H^2} + A \frac{\Delta H(H_{ext} - H_0)}{(H_{ext} - H_0)^2 + \Delta H^2}$ when the magnetic field is at 135-degree angle. Here, $\Delta H$ is the half-width at half-maximum of the resonance peak, $H_{ext}$ is the applied external field, and $H_0$ is the resonance field of the permalloy layer. The anti-symmetric resonance spikes across the field sweep seen in Fig. 5b combined with the large symmetric component of individual peaks as shown in Fig. 5c indicates strong, conventional damping like torques that are desirable for SOT-based devices.

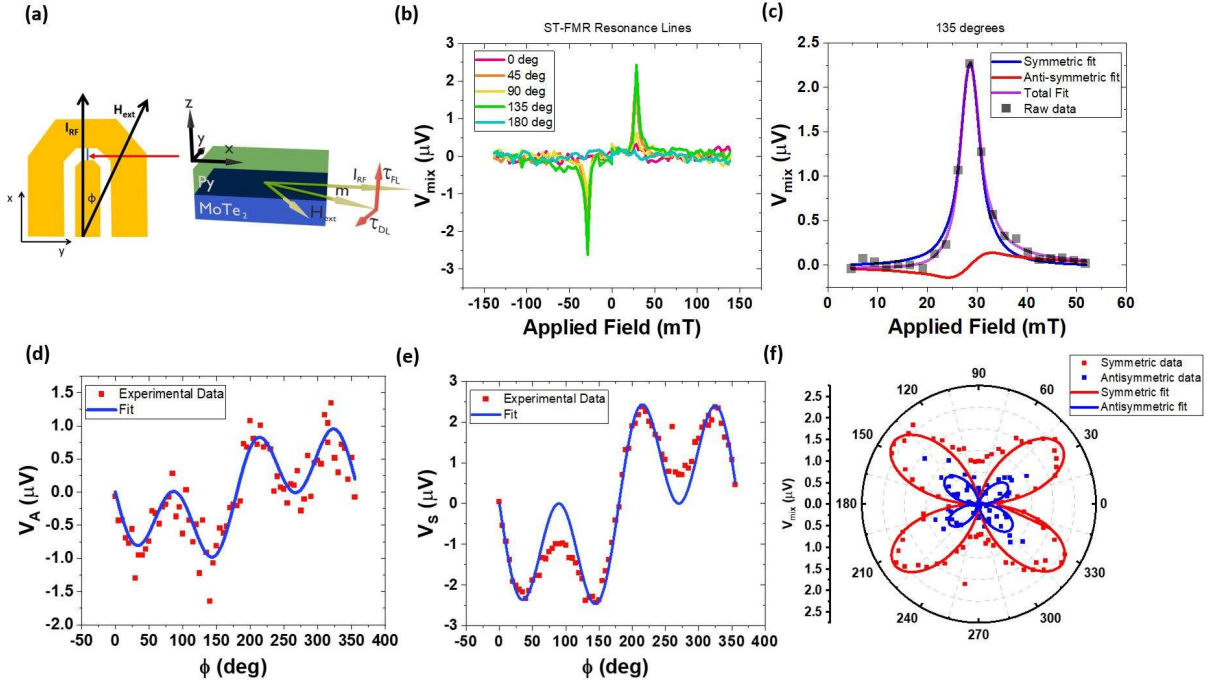

Figure 5. Spin-orbit torque (SOT) device based on 1T' MOCVD MoTe$_2$ film. (a) Schematic of the SOT based on MoTe$_2$/permalloy bilayer with a coplanar RF waveguide. (b) ST-FMR signal as a function of applied magnetic field measured at various rotation angles acquired with an RF current at 9 GHz. (c) The extracted symmetric and anti-symmetric resonance peaks from the total ST-FMR signal when the magnetic field is at 135 degree angle. (d) The extracted symmetric and (e) anti-symmetric resonance peak voltage as a function of rotation angle of the magnetic field. (f) Polar plot of the symmetric and anti-symmetric resonance peak amplitude. The symbols are measured data and lines are the fittings.

By rotating the direction of the applied field in the sample plane, we can estimate the spin-charge conversion efficiency in the bilayer. Figure 5d,e show the extracted symmetric and anti-symmetric resonance peak as a function of the magnetic field angle. At each measurement angle, the symmetric and anti-symmetric components of the resonance peaks are extracted and then fit according to the relations $V_S(\phi) = S_{DL}^x \sin(2\phi)\sin(\phi) + S_{DL}^y \sin(2\phi)\cos(\phi) + S_{FL}^z \sin(2\phi)$ and $V_A(\phi) = A_{FL}^x \sin(2\phi)\sin(\phi) + A_{FL}^y \sin(2\phi)\cos(\phi) + A_{DL}^z \sin(2\phi)$ where the $S$ and $A$ parameters represent voltage contributions from different torques for the symmetric and anti-symmetric cases, respectively, and $\phi$ is the angle between the applied field and RF current. Assuming that field like-torques in the y-direction only originate from Oersted fields, we can then estimate the spin charge conversion efficiency $\xi_{FMR}$ using the following relation: $\xi_{FMR} = \frac{S_{DL}^y}{A_{FL}^y} \frac{\sqrt{2}}{2} \frac{q\mu_0 M_s t_{FM} t_{MoTe_2}}{\hbar} \sqrt{\left(1 + \frac{4\pi M_{eff}}{H_0}\right)}$. Here, $\frac{S_{DL}^y}{A_{FL}^y}$ is the ratio between the symmetric and anti-symmetric voltage signals for the conventional torques, $M_s$ is the saturation magnetization of permalloy, $t_{FM}$ is the thickness of the permalloy layer, $t_{MoTe_2}$ is the thickness of the MoTe$_2$ film, and $M_{eff}$ is the effective magnetization calculated with the Kittel equation (determined to be approximately 0.75 T). Fig. 5f summarizes angle-dependent data for both the symmetric and anti-symmetric components. The relatively large ratio of the symmetric to anti-symmetric contributions indicates large spin-charge conversion. Using the above relation on the measured data yields an estimated $\xi_{FMR}$ of 1.66, indicating highly efficient spin-charge conversion. As $\xi_{FMR}$ also depends on interface transparency it is not necessarily equal to the spin Hall angle $\theta_{SH}$ that is typically used as a metric for spin-charge conversion potency. Accurately calculating $\theta_{SH}$ from ST-FMR data alone is challenging, though $\xi_{FMR}$ can be treated as a lower bound for $\theta_{SH}$ and be used to make rough comparisons between materials [28, 29]. Commonly cited $\theta_{SH}$ values for heavy metals are about 0.06–0.07 for platinum, 0.12–0.15 for tantalum, and 0.3 for tungsten. It has also been reported that oxidation of heavy metals can greatly enhance their $\theta_{SH}$, with values as high as 0.9 reported for oxidized platinum [20, 21, 23, 30-32]. In more exotic materials like topological insulators and Weyl semi-metals, $\theta_{SH}$ can become much larger than 1 with values up to 53 reported for materials like molecular beam epitaxy grown BiSb, though most reported values for $\theta_{SH}$ are less than 1 [29, 33]. Our estimated efficiency of 1.66 is large compared to most other materials, and our BEOL compatible growth process is more scalable compared to processes used for other large $\theta_{SH}$ materials.

## Conclusion

In conclusion, we successfully synthesized wafer-scale 1T' MoTe$_2$ using sputtered Mo and di-tert-butyl-telluride (($C_4H_9$)$_2$Te) as precursors. The synthesized 1T' MoTe$_2$ showed excellent quality, as evidenced by distinct peaks observed in Raman spectra, as well as uniformity demonstrated through EDS mapping and Raman mapping on a large scale. By adjusting various growth parameters, we demonstrated synthesis of the mixed phase 1T'/2H MoTe$_2$ using the same setup. Furthermore, the process was found to be compatible with back-end-of-line applications, with the capability of growth at 400 °C. The Hall response of the synthesized film displayed characteristic linear trends at different temperatures, and magnetoresistance of the film was also characterized. ST-FMR measurements on a 1T' MoTe$_2$ and permalloy bilayer showed strong symmetric resonance peaks, indicating the presence of a large conventional damping-like torque within the layer. We believe our findings will pave the way for wafer-scale synthesis of 1T' $MoTe_2$ using a novel tellurium precursor and offer promising prospects for the development of spintronic devices based on 1T' $MoTe_2$.

## Acknowledgement

The authors would like to acknowledge the support from Entegris Incorporated under grant No. Entegris 108252. The spin-torque ferromagnetic resonance measurements were financially supported from the Air Force Office of Scientific Research, Multidisciplinary University Research Initiatives (MURI) Program under award number FA9550-23-1-0311.

## Author information

# These authors contributed equally. Three way "first" authorship sorted alphabetically by last name.
* Corresponding email: wjzhu@illinois.edu.

## References

[1] S. Manzeli, D. Ovchinnikov, D. Pasquier, O. V. Yazyev, and A. Kis, "2D transition metal dichalcogenides," *Nature Reviews Materials,* vol. 2, no. 8, 2017, doi: 10.1038/natrevmats.2017.33.

[2] W. Choi, N. Choudhary, G. H. Han, J. Park, D. Akinwande, and Y. H. Lee, "Recent development of two-dimensional transition metal dichalcogenides and their applications," *Materials Today,* vol. 20, no. 3, pp. 116-130, 2017, doi: 10.1016/j.mattod.2016.10.002.

[3] P. Song *et al.*, "Coexistence of large conventional and planar spin Hall effect with long spin diffusion length in a low-symmetry semimetal at room temperature," *Nat Mater,* vol. 19, no. 3, pp. 292-298, Mar 2020, doi: 10.1038/s41563-019-0600-4.

[4] Q. Wang *et al.*, "Room-Temperature Nanoseconds Spin Relaxation in $WTe_2$ and $MoTe_2$ Thin Films," *Adv Sci (Weinh),* vol. 5, no. 6, p. 1700912, Jun 2018, doi: 10.1002/advs.201700912.

[5] S. Liang *et al.*, "Spin-Orbit Torque Magnetization Switching in $MoTe_2$ /Permalloy Heterostructures," *Adv Mater,* vol. 32, no. 37, p. e2002799, Sep 2020, doi: 10.1002/adma.202002799.

[6] H. Ryu *et al.*, "Laser-Induced Phase Transition and Patterning of hBN-Encapsulated $MoTe_2$," *Small,* vol. 19, no. 17, p. e2205224, Apr 2023, doi: 10.1002/smll.202205224.

[7] D. Wu *et al.*, "Phase-controlled van der Waals growth of wafer-scale 2D $MoTe_2$ layers for integrated high-sensitivity broadband infrared photodetection," *Light Sci Appl,* vol. 12, Jan 2 2023, Art no. 5, doi: 10.1038/s41377-022-01047-5.

[8] T. Kim *et al.*, "Wafer-Scale Epitaxial 1T′, 1T′-2H Mixed, and 2H Phases $MoTe_2$ Thin Films Grown by Metal-Organic Chemical Vapor Deposition," *Advanced Materials Interfaces,* vol. 5, no. 15, p. 1800439, 2018, doi: 10.1002/admi.201800439.

[9] X. Xu *et al.*, "Thermodynamics and Kinetics Synergetic Phase-Engineering of Chemical Vapor Deposition Grown Single Crystal $MoTe_2$ Nanosheets," *Crystal Growth & Design,* vol. 18, no. 5, pp. 2844-2850, 2018, doi: 10.1021/acs.cgd.7b01624.

[10] P. Zhang *et al.*, "Reduced Schottky barrier height at metal/CVD-grown MoTe$_2$ interface," *Applied Physics Letters,* vol. 120, no. 26, p. 261901, 2022, doi: 10.1063/5.0097423.

[11] S. Ha *et al.*, "Enhanced Optical Third-Harmonic Generation in Phase-Engineered MoTe$_2$ Thin Films," *ACS Photonics,* vol. 9, no. 8, pp. 2600-2606, 2022, doi: 10.1021/acsphotonics.2c00222.

[12] C. K. Safeer *et al.*, "Large Multidirectional Spin-to-Charge Conversion in Low-Symmetry Semimetal MoTe$_2$ at Room Temperature," *Nano Lett,* vol. 19, no. 12, pp. 8758-8766, Dec 11 2019, doi: 10.1021/acs.nanolett.9b03485.

[13] J. C. Park *et al.*, "Phase-Engineered Synthesis of Centimeter-Scale 1T' - and 2H-Molybdenum Ditelluride Thin Films," *ACS Nano,* vol. 9, no. 6, pp. 5627-6636, 2015.

[14] K. A. Duerloo, Y. Li, and E. J. Reed, "Structural phase transitions in two-dimensional Mo- and W-dichalcogenide monolayers," *Nat Commun,* vol. 5, p. 4214, Jul 1 2014, doi: 10.1038/ncomms5214.

[15] T. A. Empante *et al.*, "Chemical Vapor Deposition Growth of Few-Layer MoTe$_2$ in the 2H, 1T', and 1T Phases: Tunable Properties of MoTe$_2$ Films," *ACS Nano,* vol. 11, no. 1, pp. 900-905, Jan 24 2017, doi: 10.1021/acsnano.6b07499.

[16] Y. Yoo, Z. P. DeGregorio, Y. Su, S. J. Koester, and J. E. Johns, "In-Plane 2H-1T' MoTe$_2$ Homojunctions Synthesized by Flux-Controlled Phase Engineering," *Adv Mater,* vol. 29, no. 16, p. 1605461, Apr 2017, doi: 10.1002/adma.201605461.

[17] A. S. Pine and G. Dresselhaus, "Raman Spectra and Lattice Dynamics of Tellurium," *Physical Review B,* vol. 4, no. 2, pp. 356-371, 1971, doi: 10.1103/PhysRevB.4.356.

[18] F. C. Chen *et al.*, "Extremely large magnetoresistance in the type-II Weyl semimetal MoTe$_2$," *Physical Review B,* vol. 94, no. 23, p. 235154, 2016, doi: 10.1103/PhysRevB.94.235154.

[19] D. H. Keum *et al.*, "Bandgap opening in few-layered monoclinic MoTe$_2$," *Nature Physics,* vol. 11, no. 6, pp. 482-486, 2015, doi: 10.1038/nphys3314.

[20] Q. Shao *et al.*, "Roadmap of Spin–Orbit Torques," *IEEE Transactions on Magnetics,* vol. 57, no. 7, 2021, doi: 10.1109/tmag.2021.3078583.

[21] A. Hoffmann, "Spin Hall Effects in Metals," *IEEE Transactions on Magnetics,* vol. 49, no. 10, pp. 5172-5193, 2013, doi: 10.1109/tmag.2013.2262947.

[22] P. Li *et al.*, "Spin-momentum locking and spin-orbit torques in magnetic nano-heterojunctions composed of Weyl semimetal WTe$_2$," *Nat Commun,* vol. 9, no. 1, p. 3990, Sep 28 2018, doi: 10.1038/s41467-018-06518-1.

[23] L. Liu, C.-F. Pai, Y. Li, H. W. Tseng, D. C. Ralph, and R. A. Buhrman, "Spin-Torque Switching with the Giant Spin Hall Effect of Tantalum," *Science,* vol. 336, no. 6081, p. 4, 2012.

[24] W. Zhang *et al.*, "Research Update: Spin transfer torques in permalloy on monolayer MoS$_2$," *APL Materials,* vol. 4, no. 3, p. 032302, 2016, doi: 10.1063/1.4943076.

[25] T. S. Ghiasi, A. A. Kaverzin, P. J. Blah, and B. J. van Wees, "Charge-to-Spin Conversion by the Rashba-Edelstein Effect in Two-Dimensional van der Waals Heterostructures up to Room Temperature," *Nano Lett,* vol. 19, no. 9, pp. 5959-5966, Sep 11 2019, doi: 10.1021/acs.nanolett.9b01611.

[26] L. Liu, T. Moriyama, D. C. Ralph, and R. A. Buhrman, "Spin-torque ferromagnetic resonance induced by the spin Hall effect," *Phys Rev Lett,* vol. 106, no. 3, p. 036601, Jan 21 2011, doi: 10.1103/PhysRevLett.106.036601.

[27] W. L. Yang *et al.*, "Determining spin-torque efficiency in ferromagnetic metals via spin-torque ferromagnetic resonance," *Physical Review B,* vol. 101, no. 6, p. 064412, 2020, doi: 10.1103/PhysRevB.101.064412.

[28] L. Zhu, D. C. Ralph, and R. A. Buhrman, "Effective Spin-Mixing Conductance of Heavy-Metal-Ferromagnet Interfaces," *Phys Rev Lett,* vol. 123, no. 5, p. 057203, Aug 2 2019, doi: 10.1103/PhysRevLett.123.057203.

[29] A. Manchon *et al.*, "Current-induced spin-orbit torques in ferromagnetic and antiferromagnetic systems," *Reviews of Modern Physics,* vol. 91, no. 3, p. 035004, 2019, doi: 10.1103/RevModPhys.91.035004.

[30] J. C. Rojas-Sanchez *et al.*, "Spin pumping and inverse spin Hall effect in platinum: the essential role of spin-memory loss at metallic interfaces," *Phys Rev Lett,* vol. 112, no. 10, p. 106602, Mar 14 2014, doi: 10.1103/PhysRevLett.112.106602.

[31] C.-F. Pai, L. Liu, Y. Li, H. W. Tseng, D. C. Ralph, and R. A. Buhrman, "Spin transfer torque devices utilizing the giant spin Hall effect of tungsten," *Applied Physics Letters,* vol. 101, no. 12, p. 122404, 2012, doi: 10.1063/1.4753947.

[32] J. P. Fraser *et al.*, "Selective phase growth and precise-layer control in $MoTe_2$," *Communications Materials,* vol. 1, 2020, Art no. 48, doi: 10.1038/s43246-020-00048-4.

[33] N. H. D. Khang, Y. Ueda, and P. N. Hai, "A conductive topological insulator with large spin Hall effect for ultralow power spin-orbit torque switching," *Nat Mater,* vol. 17, no. 9, pp. 808-813, Sep 2018, doi: 10.1038/s41563-018-0137-y.